\def\ba{\begin{eqnarray}}
\def\ea{\end{eqnarray}}
\def\lb{\label}

\def\bi{\bibitem} 
\def\f{\frac}
\def\ts{\tilde s}

\documentclass[12pt]{article}

\usepackage{epsfig}
\usepackage{graphicx}
\begin{document} 
\title{Limiting fragmentation from scale-invariant merging of fast partons}
\author{A.Bialas and A.Bzdak\\ M.Smoluchowski Institute of Physics
 \\Jagellonian
University, Cracow\thanks{Address: Reymonta 4, 30-059 Krakow, Poland;
e-mail:bialas@th.if.uj.edu.pl}\\
Institute of Nuclear Physics Polish Academy of Sciences\thanks{Address:
Krakow, Radzikowskiego 152}\\
 R.Peschanski\\Institut de physique th{\'e}orique, CEA/Saclay\thanks{
Address:  91191 Gif-sur-Yvette cedex, France; 
e-mail:robi.peschanski@cea.fr}\ \ \thanks{URA 2306, unit\'e de recherche 
associ\'ee au CNRS.}
}
\date{}
\maketitle

PACS: 13.85.Hd, 25.75.-q, 25.75.Dw

Keywords: limiting fragmentation, self-similar parton process, scaling law

\begin{abstract}  
Exploiting the idea that the fast partons of an energetic projectile can be 
treated as sources of colour radiation interpreted as wee partons, it is shown 
that the recently observed property of extended limiting fragmentation implies a 
scaling law for the rapidity distribution of fast partons. This leads to a 
picture of  a self-similar process where, for fixed total rapidity $Y,$ the 
sources merge  with probability varying as $1/y.$
\end{abstract}

{\bf 1.} It was shown recently \cite{bb} that single particle
(pseudo)rapidity distributions in $p\!-\!p$, $d\!-\!Au$, $Cu\!-\!Cu$ and
$Au\!-\!Au$ collisions at 200 GeV c.m. energy \cite{phobos} can be
described by various superpositions of  contributions from the
``wounded" constituents of the nucleon: a quark and a diquark. The form
of the contribution from one wounded constituent $W(\eta)\approx W(y)$
was determined. Its characteristic feature is that, although peaked in
the direction of the constituent, it is not restricted to one hemisphere
but extends over almost full available phase-space \cite{bc}.

Apart from details inessential for our study, this observation shows
that particle production can be discussed independently for the
projectile and for the target. This certainly radically simplifies the
problem which, originally, may seem to be hopelessly complicated.
Moreover, it suggests that the main effect of the target on the emission
of particles from the projectile is a passive one: apparently the role
of the target is to define the phase-space available for the radiation
from the projectile\footnote{Of course the roles of projectile and target
are completely symmetric and can be exchanged one for the other.}.

The analysis of \cite{bb} was performed at one single energy. It is
interesting to investigate if (and how) the resulting picture can
accommodate the recently discovered property of {\it extended limiting
fragmentation} \cite{lf,elf} of the (pseudo)rapidity spectra of
particles produced in high-energy collisions \cite{elf,ua5}. This
{ asymptotic} property says that when plotted {\it vs.} $\xi\!=\!Y\!-\!y $ 
where $y$
is the rapidity of produced particle and $Y$ is the rapidity of the
projectile, data at { increasing energies approach a common curve (a straight 
line to a good approximation), until a certain value of $\xi$ which increases 
with increasing $Y.$}

In the present paper we discuss this problem in some detail and
formulate a model of particle emission from a strongly interacting
composite projectile which satisfies extended limiting fragmentation of
the spectra and is able to lead to a satisfactory description of data.
The model exploits the idea of distinguishing between the fast partons
treated as the sources of radiation and the wee partons considered as a
colour field emitted from the sources \cite{mcv,bj}. The emission is
described by the bremsstrahlung mechanism where the infrared cut-off,
necessary to give a physical meaning to the bremsstrahlung radiation, is
treated as a random variable.

By demanding that the observed spectra satisfy extended limiting
fragmentation, one can show that the distribution of sources must obey a
{\it scaling law}, which expresses the explicit rule how the
distribution of sources varies as a function of their rapidity $y$ for a given 
value of $Y.$ Indeed, the smaller is $y$ the less is the number of sources. As a 
consequence of  scaling, the same mechanism describes how the distribution 
evolves at fixed $y$ for varying energy of the collision. The scaling law also 
suggests that the mechanism responsible for this structure is a self-similar 
process in which the sources merge when their rapidity decreases.

In the next section the mechanism we propose is explained. The scaling
law following from extended limiting fragmentation is discussed in
Section 3. The self-similar merging  picture is  described in
Section 4. Our results are summarized and commented in the last section.

{\bf 2.} Starting with  the idea formulated in \cite{mcv}, we assume that the
wounded constituent represents a collection of partonic sources, each
one emitting soft partons which may be represented by a colour field.
Denoting the distribution of sources by $R(y_+;Y)$, where $y_+$ is the
rapidity of the source\footnote{Throughout this paper we use the
 target rest frame.} and $Y$ is the
rapidity of the projectile, we obtain for the parton distribution 
\ba
W(y;Y) =\int_y^Y dy_+\ H(y;y_+)\ R(y_+;Y)
\lb{W} 
\ea 
where $H(y;y_+)$ is
the distribution of {\it all} partons coming from a source located at
$y_+$. The final distribution ${dN}/{dy}$ is obtained from the sum of 
contributions of the wounded constituents from both particle and target sides. 

The details of $H(y;y_+)$ depend on the mechanism of radiation. 
Here, for simplicity, we take  the bremsstrahlung
or Weizsaecker-Williams distribution in its simplest form:
\ba
H(y;y_+)= \int_0^{y_+} dy_-\ P(y_-;y_+)\
 \Theta(y\!-\!y_-)\Theta(y_+\!-\!y)
\lb{flat}
\ea
i.e. a flat distribution in rapidity\footnote{The general formula for
the
distribution of partons following from the bremsstrahlung mechanism at
high energy is $a(1-x)^a+ax(1-x)^{a-1}$ with $x=e^{y-y_+}$
\cite{stod}. The first
term represents the radiated partons. The second term describes the
radiating (leading) one.
For $a=1$ we obtain the flat distribution (\ref{flat}).}, extending from
$y_+$ till some (in principle arbitrary) infra-red cut-off value $y_-$. 
$P(y_-;y_+)$ is the distribution of probability that, for a source located 
at $y_+$, the radiation is cut at $y_-$.

Physically, the very existence of the cut-off is the consequence of the
interaction with the target. Since, however, the target is a rather
complicated object, one cannot hope to evaluate explicitely the  actual
distribution of  $y_-$.
As a statistical approximation we shall assume that it can take any
 value within the
allowed kinematic limits with equal probability\footnote{We shall 
discuss some generalization at the end of the paper.}.
 This idea leads to the explicit formula for
the distribution $P(y_-;y_+)$:
\ba
P(y_-;y_+)\ dy_-= \frac1{y_+}\ \Theta(y_-)\Theta(y_+\!-\!y_-)\ dy_- \lb{p}
\ea
where the factor in front provides the proper normalization of the
probability:
\ba
\int_0^{y_+} P(y_-;y_+)\ dy_- =1\ .
\ea
All in all, we finally obtain
\ba
W(y;Y)=
y \int_y^Y\! R(y_+;Y)\ \frac{dy_+}{y_+}\ . \lb{fin}
\ea
This formula allows one to determine the distribution of produced partons
from that of the sources contained in the wounded constituent.

\vspace{.1cm}

{\bf 3.} { Let us now introduce a convenient formulation of the constraints 
imposed by the property of extended limiting fragmentation. Starting with the 
distribution $W(y,Y)$ and its derivatives $\frac{\partial^k W(y,Y)}{\partial 
y^k}$ in the target fragmentation region $y \sim Y,$ it requires on quite 
general grounds 
\ba
\frac{\partial W(y,Y)}{\partial y}\vert_{y=Y} \to cst.\ ;\quad  \frac{\partial^k 
W(y,Y)}{\partial y^k}\vert_{y=Y} \to 0, \ {\rm for}\ k\ge 2
\lb {constraints}
\ea
 when $Y$ becomes large.
Indeed, this reflects the fact that at increasing energies, the rapidity 
distribution  approaches a common curve which is a straight line. We note that 
from analyticity the higher derivatives cannot be exactly zero\footnote{The 
extended limiting fragmentation is an {\it asymptotic} property valid at large 
$Y.$ In particular it does not mean that the spectrum depends only on 
$\xi=Y\!-\!y$ even in a restricted region in $\xi$. In other terms, one has to 
decribe not only the limiting fragmentation, but also its breaking at some 
rapidity increasing with $Y.$}, but their decrease with $Y$ as expressed by the 
constraint (\ref{constraints}) implies  that limiting fragmentation region 
extends further in rapidity $Y\!-\!y$ with increasing total rapidity.

A more precise mathematical formulation of these constraints 
which   reproduces the experimental features described in the 
introduction can be summarized in the formula
\ba
W(y,Y)= (Y-y) \Phi(s)\equiv Y\Psi(s)  \label{wpsi}
\ea 
with $\Phi(s=1)\neq 0$. 
One easily sees that $W(y,Y)$ given by (\ref{wpsi}) satisfies
(\ref{constraints}).
Moreover, one sees that $W(y;Y)$ scales towards a straight line limit at 
fixed rapidity $y$ when $Y\to \infty.$ More precisely, the slope of the 
distribution $W(y;Y)$ in the limiting fragmentation region is approximately 
constant (given by $\Phi(s=1)$), with significant deviations showing up at some 
finite $s_0$ and thus 
at a value of rapidity $Y\!-\!y \sim s_0 \cdot Y$ increasing with total 
rapidity.

Using our formula (\ref{fin}), relating the particle distribution $W(y,Y)$ to 
the fast parton distribution $R(y;Y),$ 
we shall now show that (\ref{wpsi}) implies that 
 $R(y;Y)$ obeys the scaling law
\ba
R(y_+;Y)=R(y_+/Y)\equiv R(s_+) \ .\lb{scal}
\ea
Indeed, (\ref{fin}) implies 
\ba
R(y,Y)= -y\frac{\partial[W(y,Y)/y]}{\partial y}=
-s\frac{d[\Psi(s)/s]}{d s}
\ea
so that (\ref{scal}) is satisfied. }

The scaling law (\ref{scal}) determines the energy dependence of the
distribution of sources. It shows that the separation between the
sources and the field scales with the total rapidity. 
It allows to express the distribution at the
total rapidity $\omega Y$ by the distribution at the total rapidity
$Y$, i.e. the evolution of the distribution of sources: 
\ba
R(y_+;\omega Y)=R\left(\f{y_+}{\omega} ;Y\right) \equiv R\left(s^+=\f 
{y_+}{\omega Y}\right) 
\lb{scal1}. 
\ea
\begin{figure} [htb]
\includegraphics[width=6.5cm]{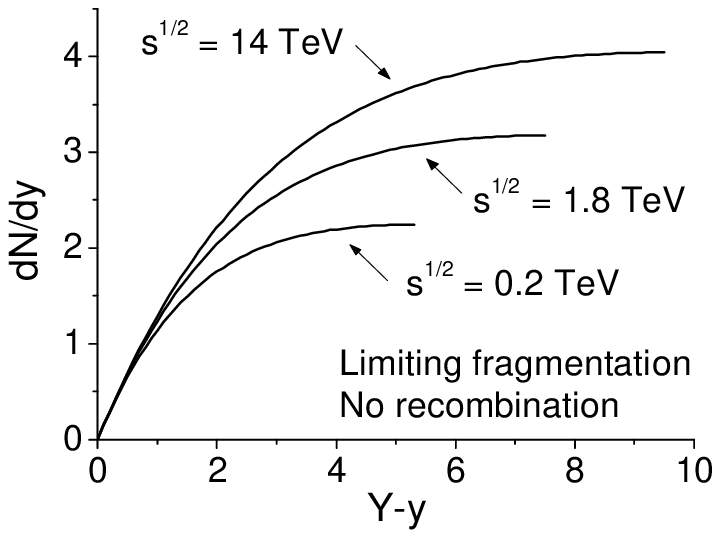}
\includegraphics[width=6.5cm]{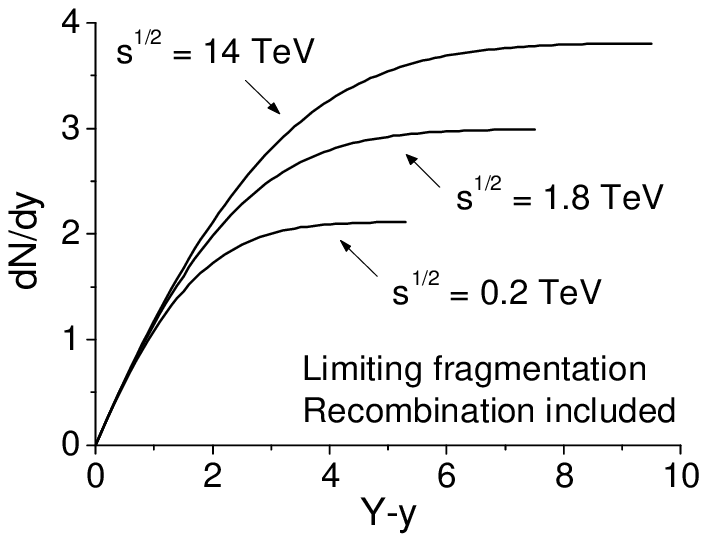}
\caption{{\it Limited fragmentation of the distribution 
of produced partons.}\newline ${dN}/{dy}$ is
 represented as a function of $Y\!-\!y$ for three  incident energies.
 Left: No recombination, $R(s_+)
\propto s_+^b,$ with $b=6.$ Right: Recombination included, $R(s_+)
\propto {s_+^b}/(1\!+\! a s_+^b)$
 with $a\! =\!3, b\! =\! 10\ .$ Note that here, by definition,  ${dN}/{dy}\equiv 
W(y;Y)\! +\!W(Y\!\!-\!\!y;Y),$ $i.e.$ the sum of contributions of constituent 
quarks from the projectile and the target.}
\lb{fig}\end{figure} This  shows explicitely how the idea of \cite{mcv} is 
realized in
our mechanism. 

The extended limiting fragmentation realized by this mechanism is shown in
Fig.1, where the particle rapidity spectra at various values of $Y$ are
plotted versus $Y\!-\!y$. One sees that  extended limiting
fragmentation is indeed well satisfied.

{\bf 4.} The scaling law (\ref{scal}) suggests that the source
distribution is formed in a self-similar cascade. Since $R$ is expected
to decrease with decreasing $s_+$, this process must correspond to
merging of fast sources into slower ones.

Indeed, introducing new variables 
$ t_+=\log y_+$, $T=\log Y $,
we see that (\ref{scal}) implies the relation
\ba
 R(y_+;Y) = R(T\!\!-\!t_+) , \lb{t}
 \ea 
and thus an increment of $T$ is equivalent to a negative shift in $t_+$.
This is typical of the cascade with equal steps and a constant
 merging probability in $t_+=\log y_+$.

In order to make the self-similar structure more explicit, let us give a 
practical example which in fact will also be physically meaningful. 

We consider a self-similar system of fast partons evolving as a function of the 
variable $t$ with  constant merging probability. For completion,
we also consider the possibility that the decrease at high density has to
be somewhat tamed if the sources screen each other. Hence, it is possible to 
write a simple equation for their  distribution, namely
\ba 
\f{dR}{dt_+} = b \left[R(t_+)-a\ R^2(t_+)\right]\ , 
\lb{merge}
 \ea 
where $b$ is the merging probability and $a$ the recombination factor. 
It is easy to write the scaling solution of (\ref{merge}) as
\ba R(y_+,Y) = \frac{R_0e^{b(t_+-T)}}{1+aR_0e^{b(t_+-T)}}=
  \f{R_0\left(y_+/Y\right)^b}{1+aR_0\left(y_+/Y\right)^b}\ . 
\lb{sol}
 \ea  
This solution  has all desired features. Indeed the source 
 distribution goes to $0$ at rapidity $y=0,$ decreases by merging when $y$ 
decreases at fixed $Y$ while, at a fixed $y$, it  decreases with 
increasing $Y$.
In fact this is the solution which was used in Fig.\ref{fig} and it is 
compatible with the phenomenological information from data, as a first 
approximation.

Finally, let us note  that such a cascade is 
 equivalent to  a  cascade with equal steps in rapidity 
  but with  the merging probability  varying as
\ba
\alpha(y)=\frac{b }{y} \ . \lb{yy}
\ea
It is interesting to observe that this result can be obtained if 
 one takes the merging probability
in the form
\ba 
\alpha (y)\ \propto \ \frac1{\log \left[d_0/d(y)\right]}
  \lb{alfa} 
\ea
where $d_0$ is the average (longitudinal) 
distance between partons in the projectile
rest frame and $d(y)$ is their average distance at the rapidity $y$.
Indeed, since the longitudinal density is basically determined by the
Lorentz factor implied by the  boost, 
one may expect that $d(y)=e^{-y} d_0$, leading
to $\alpha(y) \propto 1/y$.

One sees that (\ref{alfa}) resembles the well-known perturbative formula
for the dependence of the effective colour charge on distance. It should
be emphasized, however, that there is also an important difference
between them: while $d(y)$ and $d_0$ refer to longitudinal distances, in
the corresponding perturbative expression the four-dimensional distances
are involved. The relation between these two situations remains an
interesting problem which, however, goes beyond the scope of this paper.

{\bf 5.} Our results can be summarized as follows.

(i) We investigated the possibility that the particle production can be
described by two components: one attached to the projectile and another
one to the target both extending through most of the available
phase-space in rapidity. This idea is suggested by the success of the
phenomenological analysis based on the wounded constituent and wounded nucleon 
models \cite{bbc} which showed that such an approach can  describe 
the rapidity spectra of various processes involving hadrons and nuclei
\cite{bb,bc}\footnote{We note that a separation of this sort at high
energies has been  found in the study of nucleus-nucleus
interactions in Pomeron field theory \cite{bm}.}.

(ii) Considering one of these components (attached, say, to the
projectile) we considered the mechanism of particle production suggested
in \cite{mcv} and \cite{bj}. The partons present already in the rest
frame of the projectile are treated as sources of the bremsstrahlung
radiation producing the colour field. The distribution of sources is
determined by two effects: (a) The boost from the rest to the laboratory
frame implying a substantial increase of the parton density in
longitudinal direction, and (b) the process of recombination accompanied
by a decrease of momentum and of the density. 

(iii) The meaningful physical interpretation of bremsstrahlung radiation asks 
for the introduction of an infrared
cut-off. To implement the idea that the details of the target should not
play an important role, we assume that the cut-off is randomly
distributed through the rapidity space available for a given source. We
hope that this may be a way to take into account the 
complexity and multiplicity of elementary contributions to the soft interaction 
with the target by using  statistical arguments.

(iv) It turns out that the property of extended limiting fragmentation,
observed in all high-energy processes involving strongly interacting
particles, implies a scaling law\footnote{The existence of a scaling law
related to limited fragmentation has been already suggested in
Ref.\cite{mc}.} for the rapidity distribution of the sources. The
scaling law states that this distribution depends only on the ratio
$y/Y$ and not on $y$ and $Y$ separately. This means that the separation
between the sources and the field is not fully defined by rapidity (as
is the case in \cite{mcv}) but their proportion varies with the ratio
$y/Y$.

(v) The scaling law (\ref{scal}) suggests that the recombination
process, responsible for the distribution of sources, is a self similar
cascade. This self-similar structure differs from the cascade considered
in, e.g., \cite{mue} in few important aspects. First, it is a merging
process instead of a splitting one. Second, it is self-similar in $\log
y$ rather than in $y$. Third, the transverse momentum of partons (which
in \cite{mue} corresponds to the inverse dipole size) does not need to
be large.

\vspace{0.3cm}

Following comments are in order. 

(a) As seen from (\ref{wpsi}), extended limiting fragmentation is an
asymptotic ($Y\rightarrow \infty$) property. The rate at which it is
realized depends on the shape of the function 
\ba
\frac{s}{1-s} \int_s^1R(s_+)\frac{ds_+}{s_+}=\f{s}{1-s} 
\left[\Psi(s)-\Psi(1)\right]=
 \Phi(s)\ .
 \ea
One sees that for $R(s_+)$ varying slowly \cite{bj} near $s_+=1$
(corresponding to recombination with sizeable value of $a$ in (\ref{merge}) for 
the fast partons) one obtains a rather fast approach to the limiting line. This 
is illustrated in Fig.1. Since the data indicate that limiting
fragmentation is reached already at relatively low energies, one may
infer that indeed recombination is at work.

(b) It is worth noticing that the specific structure of the distribution of 
infra-red cut-offs as given by (\ref{p}) is not necessary to obtain the scaling 
law (\ref{scal}). Indeed, one can check that any distribution of cut-offs 
satisfying the scaling condition
\ba
P(y_-;y_+)\ dy_- = \ Q\left(\tilde s = {y_-}/{y_+}\right)\ d\ts\ 
\lb{q}
\ea
also leads to  (\ref{scal}).

(c) Our analysis concentrated on the longitudinal properties of the
spectra. It would be very interesting, of course, to elucidate also the
role of transverse momenta\footnote{In the discussion of extended limiting 
fragmentation in the framework of
CGC approach,   transverse momenta play an important role \cite{gv}.}. The key  
problem should be a better
understanding of the process of recombination of sources which may or
may not affect their transverse momentum distribution. Unfortunately, our
present approach, being purely phenomenological, does not allow to
undertake this task. 

{ d) It should be emphasized that, despite the similarity of certain
aspects of our approach with  perturbative physics, the phenomena we are 
describing are basically non-perturbative.  In particular, the scaling property 
(\ref{t}) may only imply a small {\it longitudinal} distance $d(y),$ see  
(\ref{alfa}), but not necessary  a small {\it transverse distance} ($i.e.$ large 
transverse momentum)  as for  saturation-like models (\cite{gv}).

A final comment concerns the theoretical implications of the mechanism we 
propose\footnote{See, $e.g.$ Refs.\cite{refs} for some recent theoretical 
discussions on extended limiting fragmentation.}. Indeed, this mechanism  may 
give some interesting guidelines for an exploration of non 
perturbative properties. One aspect is a theoretical hint for the   
``superposition'' property of  target and  projectile components (see $e.g.$ 
\cite{bm}). Another one is the observation that a self-similar interaction is at 
the origin the longitudinal structure of hadrons boosted to very high momenta, 
which may represent an important step in
understanding soft QCD at high energies.}

\vspace{0.3cm}
{\bf Acknowledgements}
This investigation was partly supported by the
MEiN research grant 1 P03B 045 29 (2005-2008) 
and by the VI Program of European Union "Marie Curie transfer of knowledge", 
Project: Correlations
in Complex Systems "COCOS" MTKD-CT-2004-517186. 

\end{document}